\documentclass[final,3p,times,twocolumn]{elsarticle}

\usepackage{url}
\usepackage{tabularx}
\usepackage{subcaption}
\usepackage[table,xcdraw]{xcolor}
\usepackage{siunitx}
\usepackage{multirow}
\usepackage{amsmath}
\usepackage{lineno}

\definecolor{amber(sae/ece)}{rgb}{1.0, 0.49, 0.0}

\newcommand{\revised}[1]{{\color{black}{#1}}}

\journal{Computer Methods and Programs in Biomedicine}

\begin{document}

\begin{frontmatter}

\title{gACSON software for automated segmentation and morphology analyses of myelinated axons in 3D electron microscopy}

\author[Uppsala,AIVI]{Andrea Behanova \fnref{eqC}}
\author[AIVI]{Ali Abdollahzadeh  \fnref{eqC}}
\author[Helsinki]{Ilya Belevich}
\author[Helsinki]{Eija Jokitalo}
\author[AIVI]{Alejandra Sierra\fnref{Jointsenior}\corref{cor1}} \ead{alejandra.sierralopez@uef.fi}
\author[AIVI]{Jussi Tohka \fnref{Jointsenior}\corref{cor1}} \ead{jussi.tohka@uef.fi}

\address[Uppsala]{Department of Information Technology, Uppsala University, Uppsala, Sweden.}
\address[AIVI]{Biomedical Imaging Unit, A.I.Virtanen Institute for Molecular Sciences, University of Eastern Finland, Kuopio, Finland.}
\address[Helsinki]{Electron Microscopy Unit, Institute of Biotechnology, University of Helsinki, Helsinki, Finland.}

\cortext[cor1]{Corresponding author}
\fntext[eqC]{Shared first authorship.}
\fntext[Jointsenior]{Shared senior authorship.}

\begin{abstract}
{\bf Background and Objective:} Advances in electron microscopy (EM) now allow three-dimensional (3D) imaging of hundreds of micrometers of tissue with nanometer-scale resolution, providing new opportunities to study the ultrastructure of the brain. In this work, we introduce a freely available \revised{Matlab-based} gACSON software for visualization, segmentation, assessment, and morphology analysis of myelinated axons in 3D-EM volumes of brain tissue samples. 
{\bf Methods:} The software is equipped with a graphical user interface (GUI). It automatically segments the intra-axonal space of myelinated axons and their corresponding myelin sheaths and allows manual segmentation, proofreading, and interactive correction of the segmented components. gACSON analyzes the morphology of myelinated axons, such as axonal diameter, axonal eccentricity, myelin thickness, or g-ratio. 
{\bf Results:} 
We illustrate the use of \revised{the software} by segmenting and analyzing myelinated axons in six 3D-EM volumes of rat somatosensory cortex after sham surgery or traumatic brain injury (TBI). Our results suggest that the equivalent diameter of myelinated axons in somatosensory cortex was decreased in TBI animals five months after the injury.
{\bf Conclusions:} Our results indicate that gACSON is a valuable tool  for visualization, segmentation, assessment, and morphology analysis of myelinated axons in 3D-EM volumes. \revised{It} is freely available at \url{https://github.com/AndreaBehan/g-ACSON} under the MIT license. 
 
\end{abstract}


\end{frontmatter}

\section{Introduction}
Assessing the structure of the brain is critical to better understanding its normal and abnormal functioning. Advances in electron microscopy (EM) now allow three-dimensional (3D) imaging of hundreds of micrometers of tissue with nanometer-scale resolution, providing new opportunities to study the ultrastructure of the brain \cite{Phelps2021ReconstructionMicroscopy, Nahirney2021}. Quantitative analysis of 3D-EM data, such as morphological assessment of ultrastructure, spatial distribution or connectivity of cells, requires the instance segmentation of individual ultrastructural components  \cite{Abdollahzadeh2021DeepACSONMicroscopy, Kleinnijenhuis2020AMicroscopy, Yuan2021HIVE-Net:Images}. Performing this segmentation manually is tedious, if not impossible, due to the large size and enormous number of components in typical 3D-EM data. For example, manual labeling of 215 neurites in a volume of 500 million voxels required 1500 h \cite{Berning2015}, and \revised{we estimated that manual instance segmentation of axons in a white matter EM with 300 million voxels of the size $15 \times 15 \times 50 nm^3$  would require 2400 h} \cite{Abdollahzadeh2019AutomatedMatter}. Therefore, the analyses of 3D-EM data of brain tissue require the development of advanced software tools that allow neuroscientists to automatically visualize, segment, and extract the geometric and topological features of brain ultrastructures.

There are several software tools for analyzing 3D-EM data, including open source software packages such as Microscopy Image Browser (MIB) \cite{Belevich2016}, DeepMIB \cite{Belevich2021DeepMIB:Segmentation}, Knossos \cite{Knossos}, \revised{webKnossos \cite{Boergens2017WebKnossos:Connectomics}}, AxonSeg \cite{Zaimi2016AxonSeg:Analysis}, AxonDeepSeg \cite{Zaimi2018AxonDeepSeg:Networks}, TrackEM2 \cite{Cardona2012}, CATMAID \cite{Saalfeld2009}, \revised{VAST \cite{Berger2018VASTStacks}, NeuroMorph \cite{Jorstad2018NeuroMorph:Connectivity}}, SegEM \cite{Berning2015}, Ilastik \cite{ilastik}, or MyelTracer \cite{Kaiser2021Myeltracer:Quantification}, and proprietary software such as Amira (Thermofisher Scientific$^{ TM }$, MA, \url{www.thermofisher.com/amira-avizo}) or Imaris (Oxford Instruments, Abingdon, UK, \url{www.imaris.oxinst.com}). Table \ref{tbl:software} summarizes the main features of the \revised{open source} software packages. However, the segmentation tools in these packages have mainly been designed for manual or semi-automatic segmentation, which makes their use very labor intensive for the user. The supervised learning-based segmentation methods available in these packages, as in DeepMIB \cite{Belevich2021DeepMIB:Segmentation} or AxonDeepSeg \cite{Zaimi2018AxonDeepSeg:Networks}, typically require manually annotated training data, which can be difficult and tedious to annotate. Moreover, while some of these packages have excellent visualization capabilities, they are not geared towards interactive correction, proofreading, and validation of automatic segmentations.

\begin{table*}[h!]
    \centering
    \caption{Examples of open-source software tools for 2D and 3D-EM image analysis and their main features.}

    \begin{tabularx}{\textwidth}{lX}
    {\cellcolor[rgb]{0.937,0.937,0.937}}\textbf{Software} & {\cellcolor[rgb]{0.937,0.937,0.937}}\textbf{Main features}\\ \hline\hline
    
    \multirow{3}{*}{MIB \cite{Belevich2016}} & Matlab-based; manual and semi-automated segmentation of large EM volumes (graph cuts method, supervoxels); \revised{reading and writing a variety of image formats; many image processing tools; excellent alignment features.} \\ \hline
    
    \multirow{3}{*}{DeepMIB \cite{Belevich2021DeepMIB:Segmentation}} & Matlab-based; supervised deep learning-based image segmentation of EM and light microscopy volumes; \revised{includes all MIB image processing tools; user friendly GUI for training and inference of deep models.} \\ \hline
    
    \multirow{3}{*}{Knossos \cite{Knossos}} & Python-based; \revised{3D visualization of large EM volumes, not limited to the RAM size; visualizing large EM volumes; skeletonization by manually placing and connecting nodes; manual processing of pre-segmented data.} \\ \hline
    
    \multirow{3}{*}{webKnossos \cite{Boergens2017WebKnossos:Connectomics}} & In-browser; \revised{fast browsing speeds; efficient distributed EM annotation in petabytes-scale; provides flight mode, a single-view egocentric reconstruction method to accelerate annotation.} \\ \hline
    
    \multirow{3}{*}{AxonSeg \cite{Zaimi2016AxonSeg:Analysis}} & Matlab-based; automated 2D segmentation of axons and myelin on histology images (the extended minima algorithm; active contours to detect the outer and the inner boundaries of myelin sheath); \revised{extracts morphometric information, e.g., axon diameter or myelin g-ratio.} \\ \hline
    
    \multirow{3}{*}{AxonDeepSeg \cite{Zaimi2018AxonDeepSeg:Networks}} & Python-based; supervised deep learning-based software for semantic segmentation of axons and myelin from microscopy images (3-class semantic segmentation task); \revised{easy training procedure.}  \\ \hline    
    
    \multirow{3}{*}{TrackEM2 \cite{Cardona2012}} & ImageJ plugin; manual and semi-automated segmentation of large EM volumes (brushing, skeletonization); \revised{image stitching, registration (using scale invariant feature transform), editing, and annotation features (floating text labels).} \\ \hline 
    
    \multirow{3}{*}{CATMAID \cite{Saalfeld2009}} & Python-based; fast terabyte-scale image data browsing; user-interactive image annotation; WebGL neuronal morphology viewer; \revised{multiple linked image stack display; many tools for collaborative microcircuit reconstruction} \\ \hline
    
    \multirow{3}{*}{VAST \cite{Berger2018VASTStacks}} & C++ \revised{based with Direct3D graphics for optimal performance; petabyte-scale image data browsing; many general labeling tool; annotating arbitrary structures, regions and locations, depending on the user’s needs; proof-reading of segmentations.} \\ \hline 
    
    \multirow{3}{*}{NeuroMorph \cite{Jorstad2018NeuroMorph:Connectivity}} & Operating \revised{in Blender; import and visualize mesh models; cross-sectional analysis of axons and their synaptic connections; measuring density of organelles; mesh objects should be generated by other 3D image segmentation software.} \\ \hline

    \multirow{2}{*}{SegEM \cite{Berning2015}} & Matlab-based; semi-automated segmentation of 3D-EM data using convolutional neural networks and manually provided neuronal skeletons. \\ \hline    
    
    \multirow{3}{*}{Ilastik \cite{ilastik}} & Python-based; automated and semi-automated segmentation of EM volumes (graph cut methods for instance segmentation); \revised{object tracking; interactive processing of image sub-volumes, followed by offline batch mode for complete volume analysis.} \\ \hline    
    
    \multirow{3}{*}{MyelTracer \cite{Kaiser2021Myeltracer:Quantification}} & Python-based; 2D semi-automated segmentation of axons and myelin (intensity thresholding and user's drawn lines to generate contours for myelin segmentation); \revised{quantification of myelin g-Ratio.} \\ \hline  
    
    \multirow{3}{*}{gACSON} & Matlab-based; \revised{automated segmentation and cross-sectional morphology analysis of myelinated axons in EM volumes; instance segmentation of myelin; myelin g-ratio measurement; 3D volume visualization, provides proof-reading tools.} \\ \hline
    
    \label{tbl:software}    
    \end{tabularx}
\end{table*}

In this context, we developed a \textbf{G}raphical user interface for \textbf{A}utomati\textbf{C} \textbf{S}egmentation of ax\textbf{ON}s, called gACSON, and axonal morphology analysis in large 3D-EM data of the brain. The \revised{Matlab-based} gACSON builds on our previous pipelines ACSON \cite{Abdollahzadeh2019AutomatedMatter} and DeepACSON \cite{Abdollahzadeh2021DeepACSONMicroscopy} for 3D-EM segmentation and morphometrics and provides an easy-to-use graphical user interface (GUI) and additional segmentation algorithms for myelin and the intra-axonal space of myelinated axons. The software allows for automatic unsupervised segmentation that requires no manually annotated training data. While ACSON and DeepACSON are automated segmentation pipelines, gACSON is an interactive tool, providing user-friendly tools to review and correct the  segmentations. \revised{It} also allows segmentation of a subset of all myelinated axons so that the user does not need to wait for the completion of the segmentation of all myelinated axons. More importantly, \revised{the software} provides instance segmentation of myelin that maps myelin sheaths to the appropriate intra-axonal space instance. This feature makes \revised{the new tool} unique compared to our existing ACSON and DeepACSON pipelines \cite{Abdollahzadeh2019AutomatedMatter}. With the software, users can visualize and review the segmented ultrastructure or correct the results of an automated segmentation by providing a dynamic environment for interactive merge or split operations on the segmented components. Like the ACSON and DeepACSON pipelines, gACSON extracts morphological parameters of segmented myelinated axons, such as axonal diameter or eccentricity, but extends the morphological parameters to measure myelin thickness and g-ratio by enabling instance segmentation of myelin. Moreover, visualization, segmentation evaluation, or correction of automatic segmentation results is not limited to EM but can be applied to other biomedical imaging modalities, such light microscopy images. 

\begin{figure*}[t]
    \centering
    \includegraphics[width=1\textwidth]{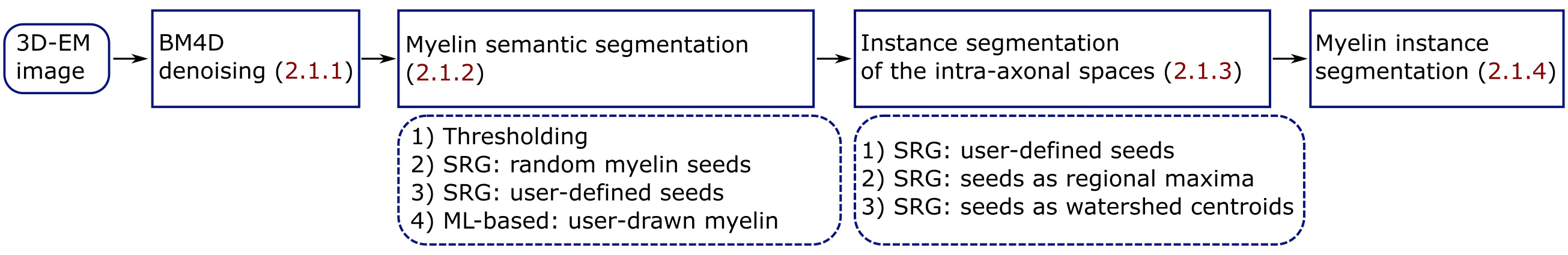}
    \caption{gACSON segmentation pipeline. The pipeline takes 3D-EM volume as input and 1) denoises the EM volume using the BM4D method, 2) segments the myelin in the denoised volume using a user-defined method (thresholding, seeded region growing (SRG) or an ML-based approach), 3) performs the instance segmentation of the intra-axonal space of myelinated axons, and 4) performs the instance segmentation of myelin. The numbers in parentheses refer to subsections where the steps are detailed.}
    \label{pipeline}
\end{figure*}

The structure of this paper is as follows: Sections \ref{seg_pip} and \ref{morph_pip} present the gACSON segmentation and morphometry pipelines, respectively. Section \ref{visualization} introduces the GUI including the parameter settings and the tools for proofreading and correcting the automatic segmentation and evaluating the accuracy of the segmentation results. In Section 3, we demonstrate the application of the software to segment a new dataset of the six serial block-face scanning electron microscopy (SBEM) volumes of rat somatosensory cortex. We further make these volumes and segmentations publicly available. 

\section{Methods}
\subsection{gACSON segmentation pipeline} \label{seg_pip}
gACSON includes an automated segmentation pipeline that traces myelinated axons in 3D-EM data of brain tissue. The segmentation pipeline, summarized in Figure \ref{pipeline}, consists of following steps: 1) denoising of EM volumes by the block-matching and 4D filtering (BM4D) algorithm \cite{bm4D} (Section \ref{bm4d}), 2) semantic segmentation of myelin (Section \ref{myelin_semantic}), 3) instance segmentation that provides a separate label to each myelinated axon (Section \ref{axon_instance}), and 4)  assignment of myelin to individual intra-axonal spaces, referred to as instance segmentation of myelin (Section \ref{myelin_instance}). \revised{All the parameters of gACSON are summarized in Supplementary Table 1.}

\subsubsection{BM4D denoising} \label{bm4d}
\begin{figure}[t]
    \centering
    \includegraphics[width=0.95\columnwidth]{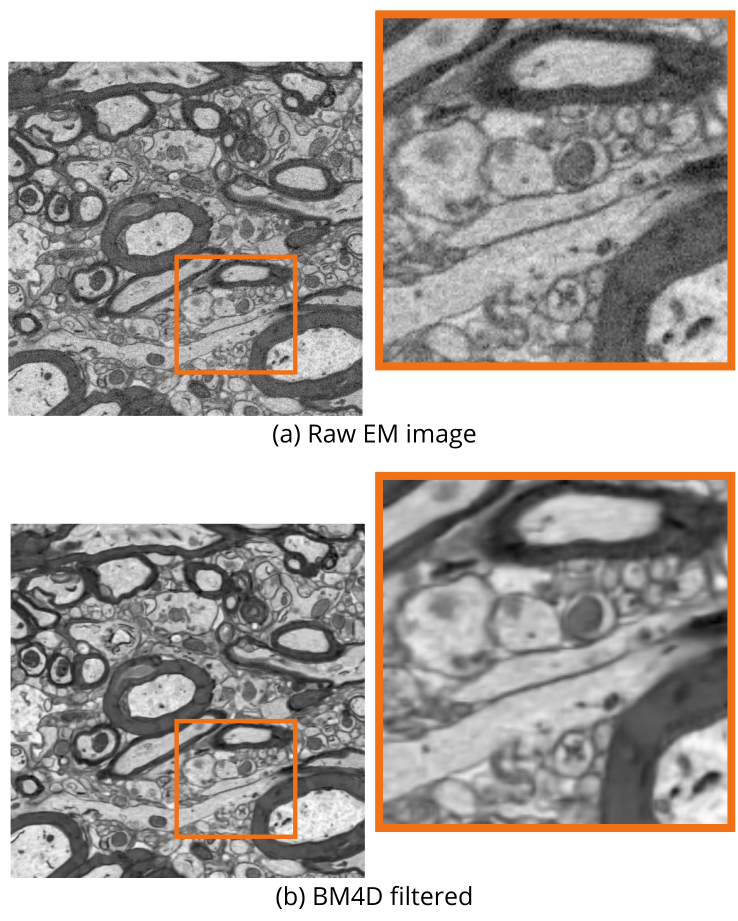}
    \caption{\textbf{a} A representative SBEM image from the somatosensory cortex of a sham-operated rat. (\textbf{b}) BM4D denoising of \textbf{a}. \revised{The magnification factor of the zoomed-in images in the orange boxes is 300 \%.}}
    \label{bm4d_filt}
\end{figure}

SBEM images are affected by primary beam noise, secondary emission noise, and noise in the final detection system \cite{Sim2006a}. \revised{We suppress these complex noise patterns because the segmentation algorithms in steps 2 and 3 are affected by it. Specifically, region growing algorithms are sensitive to image noise as a voxel belonging to a region may seem dissimilar from the average intensity of the region simply because of noise. For denoising, we apply the non-local BM4D algorithm \cite{bm4D}. Unlike local denoising filters operating in a small window around the target voxel, non-local denoising considers all the voxels in the image and weights them by the similarity of their neighbourhood to the neighbourhood of the target voxel. This results in better preservation of the image texture and edges and reduces the loss of details in the image compared with local denoising algorithms.} BM4D, \revised{in particular}, enhances a sparse representation in the transform domain by grouping similar 3D image patches (i.e., continuous 3D blocks of voxels) into 4D data arrays. BM4D automatically estimates noise type and variance and we use BM4D's default parameter values. Figure \ref{bm4d_filt} illustrates a representative SBEM image from the somatosensory cortex of a sham-operated rat and its BM4D-denoised version.

\subsubsection{Semantic segmentation of myelin} \label{myelin_semantic}
\begin{figure}[t]
    \centering
    \includegraphics[width=0.47\textwidth]{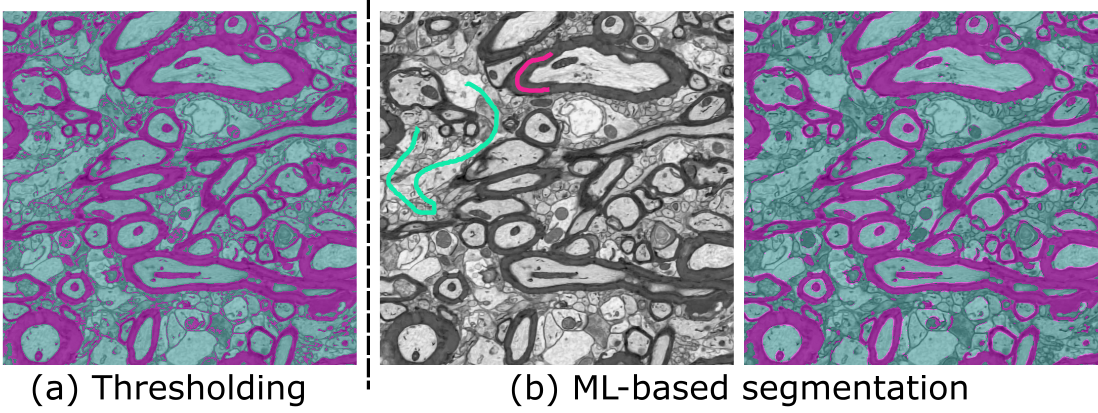}
    \caption{(\textbf{a}) Semantic segmentation of myelin by a user-defined threshold. (\textbf{b}) Machine learning (ML)-based semantic segmentation of myelin. Left panel: user specifies myelin (magenta) against other ultrastructures (green) by drawing a scribble on an EM image. These scribbles provide data to train a random forest model. Right panel: the semantic segmentation of myelin (magenta) against other ultrastructures (green). The thresholding-based segmentation of myelin incorrectly includes cell membranes and is therefore \revised{noisier} than the ML-based segmentation. Both methods segment mitochondria as myelin.}
    \label{my_semantic}
\end{figure}

gACSON is equipped with four different methods for semantic segmentation of myelin: 1) interactive segmentation of myelin by intensity thresholding (Figure \ref{my_semantic}a); 2) use of seeded region growing initiating with an automatic seed selection; 3) use of seeded region growing with user-defined seeds (manual seed selection); and 4) use of weakly supervised machine learning (ML) by training a random forest \cite{Breiman2001} with a scribble annotation (see \cite{lee2010image}) drawn by the user on \revised{2D slices} of 3D-EM data. The user can perform semantic segmentation of myelin using various algorithms that enable accurate, automated, or computationally efficient segmentation. For example, a ML -based segmentation produces a more accurate segmentation, but requires user input and can be more computationally intensive than a thresholding approach. 

In ML-based segmentation, \revised{we} use the following features to classify voxels: 1) the image intensity, 2) the Gaussian-smoothed image intensity ($\sigma = 2)$, 3) the Laplacian of the Gaussian image ($\sigma = 0.5$), 4) the Gaussian gradient magnitude image ($\sigma = 2$), and 5) the difference of Gaussians ($\sigma = 5$). $\sigma$ is the standard deviation of the isotropic Gaussian smoothing kernel in voxels. Figure \ref{my_semantic}b-left panel shows an annotation example drawn by the user to segment myelin against other components, and Figure \ref{my_semantic}b-right panel is the resulting semantic segmentation.
\revised{Scribbles can be drawn on single or multiple 2D slices of the 3D volume leading to a training set to train the random forest using the above described features for each voxel. The trained random forest then automatically extends training set to predict the myelin class in the whole 3D EM volume. The user can scroll between the 2D images and add new scribbles if needed. It is not necessary to adapt means or contrast of the image, as we assume that contrast normalization has been applied as a pre-processing step to 2D images of the 3D-EM volume.} 

Regarding seeded region growing, gACSON automatically selects the seeds for seeded region growing by either randomly distributing the seeds over the image region or placing seeds based on the intensity values in the images.

\subsubsection{Instance segmentation of the intra-axonal space of myelinated axons} \label{axon_instance}
\begin{figure}[t]
    \centering
    \begin{subfigure}[t]{0.47\textwidth}
    \includegraphics[width=\textwidth]{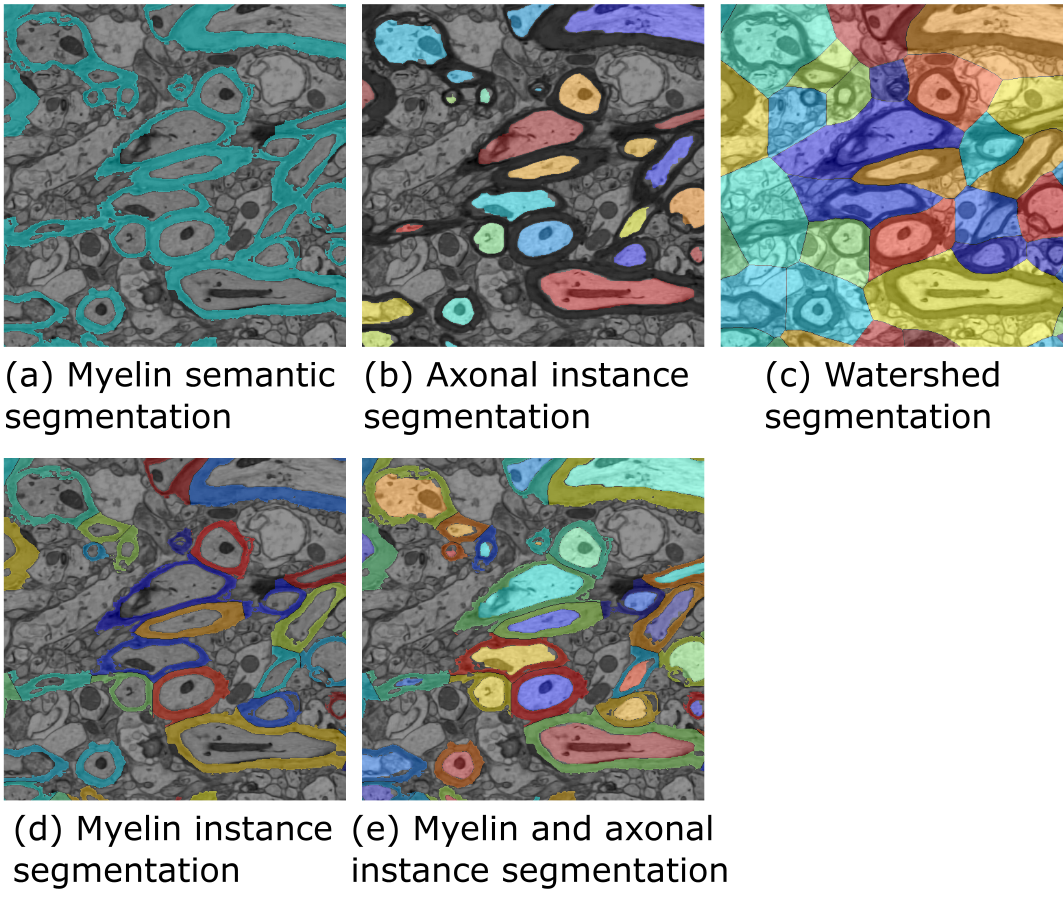}
    \end{subfigure}
    \caption{The workflow of instance segmentation of myelin. (\textbf{a}) Semantic segmentation of myelin. (\textbf{b}) Instance segmentation of the intra-axonal space of myelinated axons. (\textbf{c}) Watershed transformation using the segmented intra-axonal spaces partitions the image space into disjoint subvolumes. (\textbf{d}) Instance segmentation of myelin. (\textbf{e}) Instance segmentation of myelin and the intra-axonal spaces. Colors of distinct instances in the figure have been randomly generated and they are not comparable between different panels.}
    \label{my_instance}
\end{figure}

This step segments the complement of the myelin segment \revised{(see Figure \ref{my_instance}a for an example of semantic segmentation of myelin)} into instances of myelinated axons using the Bounded Volume Growing algorithm (BVG) (\revised{See  Figure \ref{my_instance}b for an example of instance segmentation of the intra-axonal space)}. \revised{The BVG} algorithm, that we presented in \cite{Abdollahzadeh2019AutomatedMatter}, integrates seeded region growing \cite{RegionGrowing} and the Canny edge detector \cite{Canny1986ADetection} to segment the intra-axonal space of myelinated axons. BVG iteratively segments an EM volume into myelinated axons $V_1,\ldots, V_N$, where $V_0$ denotes the myelin segment. Each myelinated axon $V_i$ is segmented based on region growing starting from a seed voxel and adding more voxels from the neighborhood of the current $V_i$ to $V_i$ until no more voxels can be added. For a voxel to be added to $V_i$, it must meet three criteria (the voxel must not be an edge voxel; the voxel must not belong to any $V_j$ with $j < i$; the voxel must have an intensity similar to the average intensity of $V_i$). BVG grows one seed at a time and then moves on to the next seed. At the completion of an axon, all remaining seeds within the segmented axon are discarded. For a detailed technical description of the algorithm, see \cite{Abdollahzadeh2019AutomatedMatter}. The key to successful segmentation by BVG is the selection of seed voxels. gACSON allows the user to choose between three seed generation techniques:
\begin{enumerate}
\item  Initial seeds by the regional maxima method uses the semantic segmentation of the myelin to define the positions of the seeds as the myelin encloses the axons. \revised{In more detail, we compute} the Euclidean distance transform of the semantic segmentation of the myelin for each 2D image slice along the $z$ axis \revised{and generate} the seeds by extracting the regional maxima from the distance transform images. 

\item Initial seeds as the watershed centroids: the regional maxima method for generating seeds may overestimate the number of initial seeds, which increases the computation time. To save computation time, \revised{this seeding method} uses the H-maxima transform of the distance transform of myelin segmentation as a marker for a marker-controlled watershed segmentation \cite{Najman1994} and defines the centroids of the watershed segments as starting seeds for the BVG algorithm. The H-maxima transform suppresses all local maxima whose height is less than $H$. The marker-controlled watershed transform divides the image into regions one-to-one based on the positions of the markers, such that the number of markers equals the final number of watershed regions. 
\item Manual placement of start seeds: the user can manually select start seeds for the bounded volume growing algorithm. Manual seed selection is preferred for fast segmentation of a limited number of axons.
\end{enumerate} 
The accuracy, automation level, and computational efficiency of a segmentation are determined by the choice of seeding method. For example, defining seed positions using watershed centroids is computationally more efficient than defining them using regional maxima because the number of seeds is reduced. However, for the same reason, an axon is more likely to remain unsegmented when watershed centroids seeds are used.

After the initial segmentation of the intra-axonal space of myelinated axons using BVG, gACSON refines the segmentation of large volumes $V_i$ (with more than $5000$ voxels) using supervoxels generated by the simple linear iterative clustering (SLIC) method \cite{achanta2012slic}. In more detail, suppose that we have $Q$ supervoxels $S_q,\; q = 1,\ldots,Q$. Then, we re-define $V_i$ as $V_i^\prime = \bigcup \limits_{q \in I_i} S_q $ where 
$$
I_i = \bigg\{q^\prime \bigg| \dfrac{|S_{q^\prime} \cap V_i|}{|S_{q^\prime}|} \geq 0.8 \land \forall j < i: q^\prime \notin I_j  \bigg\}.
$$ 
The second condition states that a supervoxel can belong to only a single axon. The volumes larger than 5000 voxels are traversed in the order they are indexed in the segmentation.  Refining the segmentation with SLIC technique eliminates most of the small volumes by relabeling and attaching them to the larger segments. The segmentation of myelinated axons may include unmyelinated axons or fragments of other cell processes. To decide whether an axon is myelinated or unmyelinated, \revised{the method} investigates the thick hollow cylinder enclosing the axon of interest. The enclosing cylinder is formed by those supervoxels having a common face with the axon. If the enclosing cylinder contains more myelin than a threshold value, e.g., when more than $70\%$ of the cylinder volume is myelin, then the axon is considered as myelinated. 

\begin{figure}[t]
    \centering
    \begin{subfigure}[t]{0.47\textwidth}
    \includegraphics[width=\textwidth]{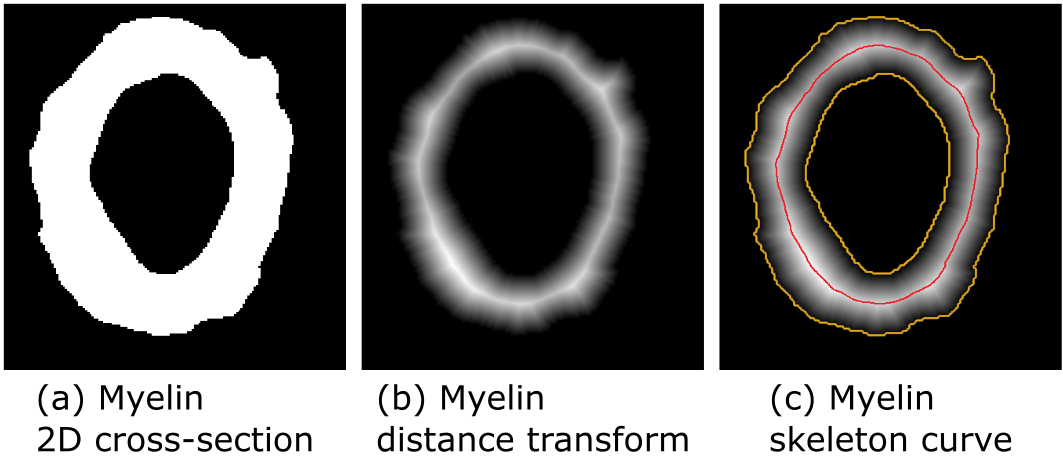}
    \end{subfigure}
    \caption{Measurement of myelin thickness. (\textbf{a}) A cross section of a segmented myelin. (\textbf{b})The Euclidean distance transformation of segmented myelin. (\textbf{c}) The myelin curve skeleton (red) and the boundaries of the myelin (yellow) are superimposed on \textbf{b}. The myelin curve  skeleton remains in the middle of the myelin cross section. The value of the distance transform of the myelin cross-section at any point on the myelin skeleton curve is the half-thickness of the myelin at that point.}
    \label{my_thick}
\end{figure}

\subsubsection{Instance segmentation of myelin} \label{myelin_instance}
gACSON provides instance segmentation of myelin by mapping myelin voxels to the corresponding intra-axonal space, with a one-to-one correspondence between myelinated axons and myelin instances, see Fig. \ref{my_instance}. To this end, \revised{we compute} the distance transform of the complement of the union of segmented intra-axonal spaces, followed by the watershed transform. Figure \ref{my_instance}c shows how the watershed transform divides the image space into $n$ disjoint subvolumes, where $n$ is the number of intra-axonal spaces. To achieve the instance segmentation of the myelin, \revised{we assign} the labels from the watershed segmentation of the intra-axonal spaces to the myelin voxels, as shown in Figure \ref{my_instance}d, and assigns the same label to the myelin voxels as to the corresponding intra-axonal space, as shown in Figure \ref{my_instance}e. Note that the watershed-based segmentation divides myelin content equally between two adjacent axons.

\begin{figure*}[t]
\centering
\includegraphics[width=\linewidth]{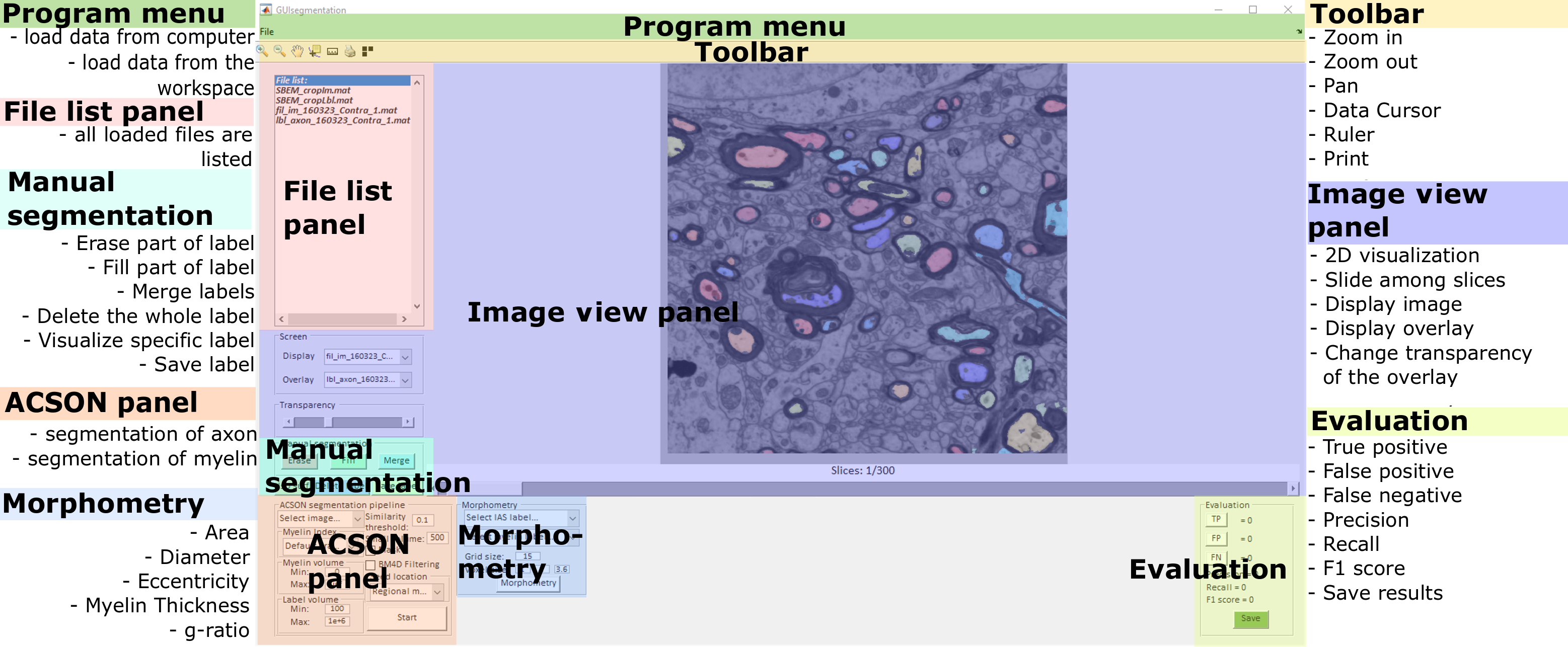}
\caption{The user-interface of the gACSON software.}
\label{GUI}
\end{figure*}

\subsection{gACSON morphometry analysis} \label{morph_pip}
The gACSON morphometry algorithm extracts and quantifies morphological features such as equivalent diameter, eccentricity, or g-ratio of segmented myelinated axons. \revised{The algorithm} automatically finds the axonal skeleton based on a distance transformation-based algorithm as described in \cite{Hassouna2005RobustSets, Abdollahzadeh2021CylindricalObjects} and extracts the cross-sections of the axons using a perpendicular plane to the axonal skeleton at each point along the axonal skeletons. For an EM volume with an anisotropic voxel size, \revised{we} first resample the bounding box of an axon to an isotropic voxel size,  and then quantifies the morphology parameter on the resampled volume. For each 2D cross-section, \revised{the software} quantifies the equivalent diameter of the intra-axonal space, i.e., the diameter of a circle with the same area as the cross-section, the length of the minor and major axes of a fitted ellipse, and its eccentricity.

\revised{The morphometry algorithm} extracts the myelin thickness and g-ratio for each myelinated axon. To measure the myelin thickness in a cross-section (Figure \ref{my_thick}a), \revised{it} first computes the Euclidean distance transform of the myelin cross-section, as shown in (Figure \ref{my_thick}b), where higher intensity values imply longer distances to the myelin boundaries. \revised{Thereafter, it} calculates the curve that remains in the center of the myelin cross-section, called the myelin skeleton curve, using the Euclidean distance transform map (Figure \ref{my_thick}c). The value of the distance transform at each point on the myelin skeleton curve is half the myelin thickness at that point. 
We use the median of the myelin thickness values to represent the myelin thickness for a cross-section of the myelin instance.  The g-ratio of the myelinated axon \revised{is calculated} as $g = \frac{d_e}{d_e + d_m}$, where $d_e$ is the equivalent diameter of the intra-axonal space and $d_m$ is the myelin thickness.

\subsection{Graphical User Interface} \label{visualization}
The gACSON user interface is shown in Figure \ref{GUI}. The \textit{program menu} panel is used to load image files or import images from the Matlab workspace using \textit{file} menu. Images from the \textit{file list} panel are displayed using the \textit{display} popup menu of the \textit{image view} panel. The user can select the \textit{overlay} pop-up menu to visualize an overlay image, such as a segmentation overlaid with the corresponding intensity image, or change the transparency of the overlay image by moving the \textit{Transparency} slider. The toolbar contains buttons to zoom in/out or display the value of a selected voxel. The toolbar's \textit{Ruler} button measures the Euclidean distance between any two points selected by the user on the displayed image. The toolbar also contains a multiplanar view for 3D image files. The software displays 2D slices, in the \textit{x-y} plane, of 3D-EM data, allowing the user to browse between 2D slices along the \textit{z}-axis.

\subsubsection{ACSON segmentation panel} \label{param_set}
The \textit{ACSON segmentation} panel sets the parameter values of the gACSON pipeline to segment the myelin and intra-axonal space of myelinated axons in 3D- EM data, as described in Section \ref{seg_pip}. 

\subsubsection{Manual segmentation panel} \label{manual_seg}
\revised{The \textit{manual segmentation} panel allows} the user to manually segment images and correct the results of an automatic segmentation method, \revised{e.g., a leakage at nodes of Ranvier or if staining artifacts were mistakenly segmented as myelin.} The user can manually draw a 2D region to either create a new label or append the region to an existing label, draw a 2D region to subtract from an existing label, e.g., to correct segmentation boundaries. \revised{The user can also discard a segmentation label, for example, when a myelinated axon is partially outside the image domain, and replace it with a null label representing the background}. The user can merge multiple components into a single label to correct the over-segmentation error, i.e., a single object segmented into multiple components. It is also possible to display the 3D representation of the surface of a segment selected by the user, which allows 3D evaluation of the shape of the segment to detect possible under- or over-segmentation errors.   

\subsubsection{Evaluation panel} \label{evaluation}
\begin{figure}[t]
\centering
\includegraphics[width=\linewidth]{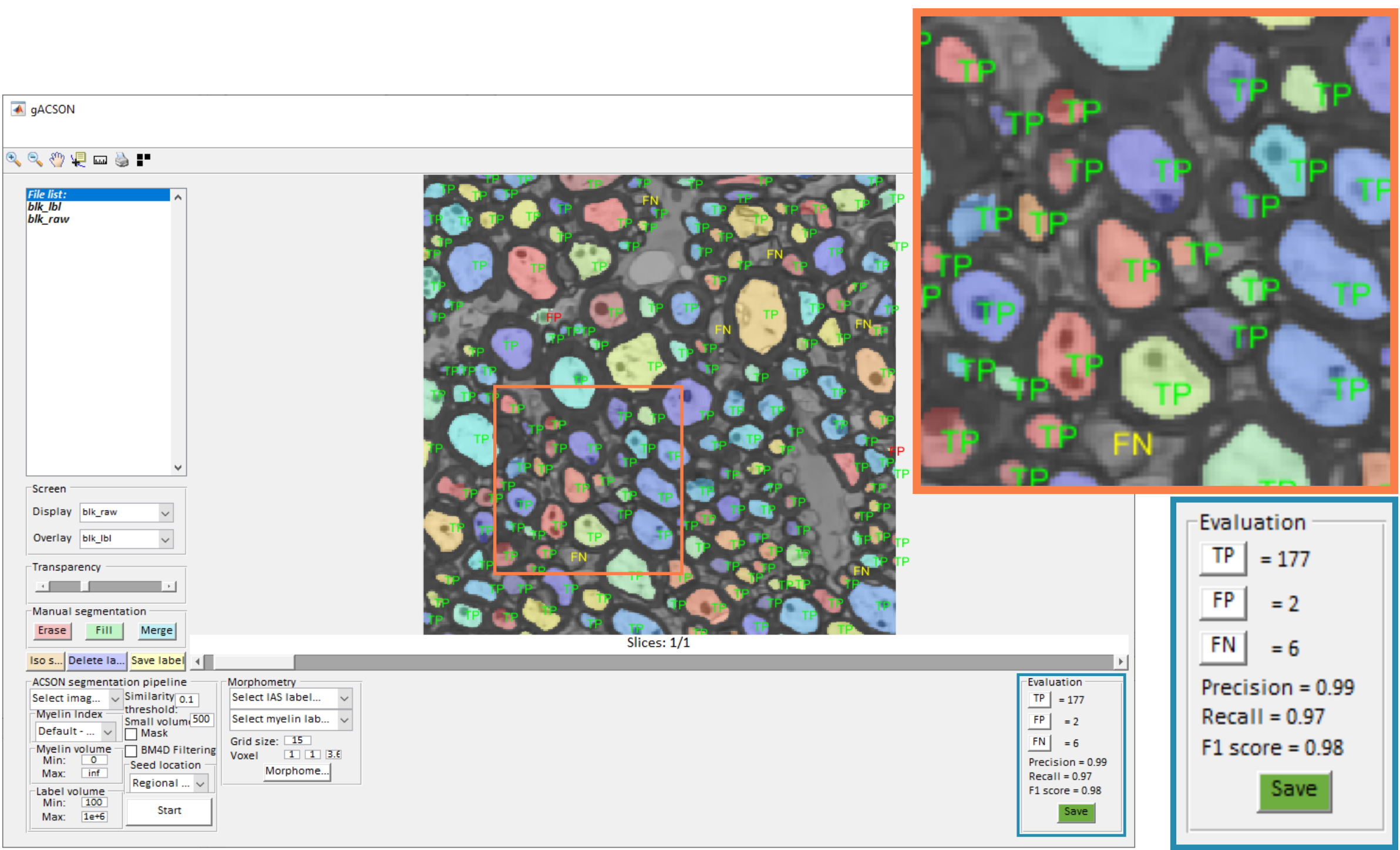}
\caption{gACSON evaluation panel to assess a segmentation. The expert evaluates the segmentation at the object level, quantifying the number of true-positives (TP), false-positives (FP), and false-negatives (FN) to calculate the precision, recall, and F1 score. We have applied the evaluation panel of gACSON to validate our DeepACSON pipeline for the instance segmentation of axons in corpus callosum and cingulum in \cite{Abdollahzadeh2021DeepACSONMicroscopy}.}
\label{eval_panel}
\end{figure}

The \textit{evaluation} panel quantifies the accuracy of instance segmentation at the object-level using precision (positive predictive value), recall (sensitivity), and F1 score (harmonic mean of precision and recall measures).  To calculate these metrics, the user manually marks all true positives (correctly detected segments), false positives (incorrectly detected segments), and false negatives (undetected segments) on a cross-section of a segmentation overlaid on the EM image. This allows users to assess the validity of a segmentation as a whole without a need for manually segmenting complete images which would be burdensome and time consuming. Figure \ref{eval_panel} demonstrates an expert \revised{evaluating} the performance of an EM segmentation algorithm.

\subsubsection{Morphometry panel} \label{morph_param_set}
\revised{The morphometry panel is used} to set the parameter values of the proposed morphometry algorithm of Section \ref{morph_pip}. \revised{The morphometry algorithm}  requires the 3D instance segmentation of myelinated axons, both the intra-axonal space and myelin, to calculate the true cross-sectional morphology of myelinated axons in 3D. The length of the minor and major axes of a fitted ellipse, equivalent diameter, eccentricity, myelin thickness, and g-ratio are measured for each axonal cross-section.

\begin{figure}[t]
\centering
\includegraphics[width=0.9\linewidth]{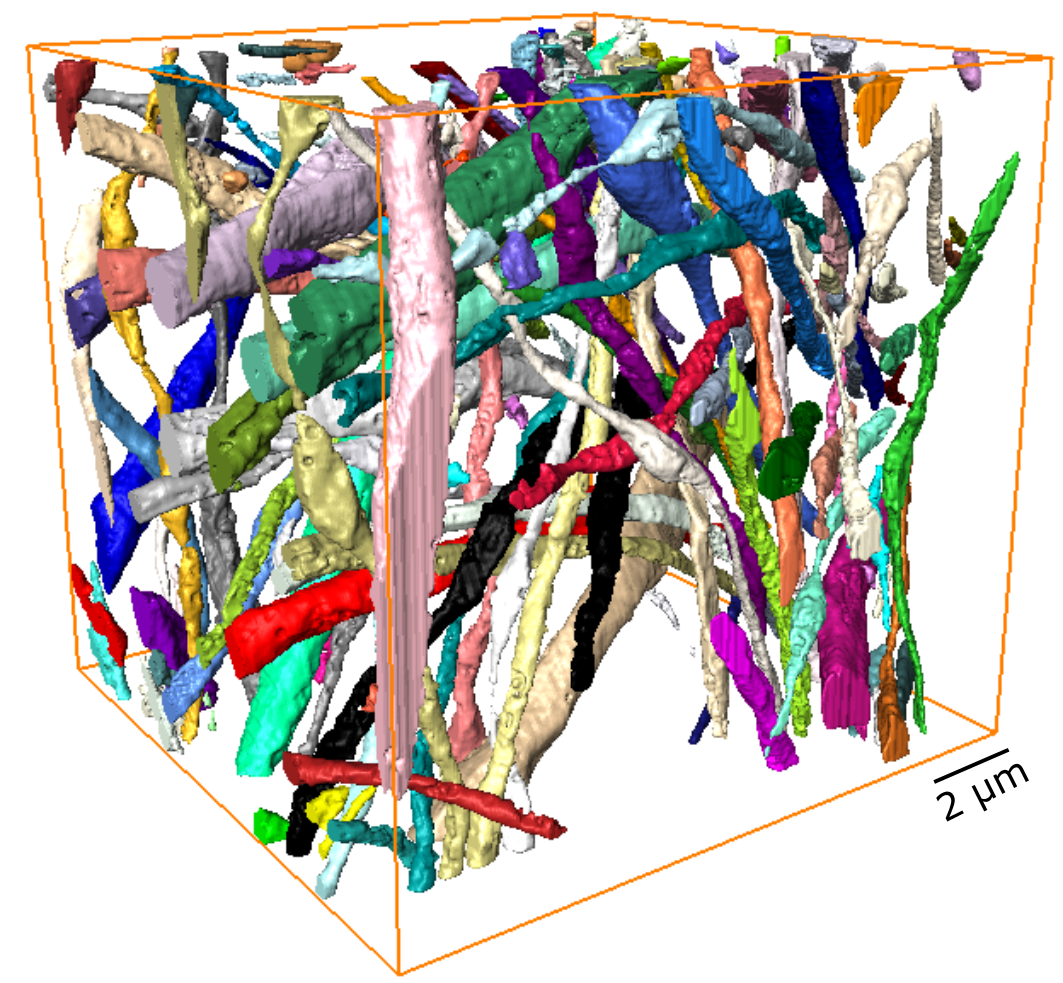}
\caption{3D representation of the intra-axonal space of myelinated axons in the somatosensory cortex of a TBI animal. The field of view of the segmented volume is $14 \times 14 \times 15$ \SI{}{\cubic \micro\meter}, which corresponds to $1024 \times 1024 \times 300$ voxels in $x$, $y$, and $z$ directions, respectively. Note the variability of axonal diameter between myelinated axons.}
\label{seg_volume}
\end{figure}

\begin{table*}[t]
    \centering
    \caption{The size and resolution of the SBEM volumes. The column Axons provides the number of myelinated axons segmented by gACSON that were longer than \SI{4}{\micro\meter}. }
    
    \begin{tabular}{l|lllll}
    \rowcolor[rgb]{0.937,0.937,0.937} \textbf{ Rat } & \textbf{Hemisphere } & \textbf{Volume ($\mu m^3$) } & \textbf{Resolution ($nm^3$) } & \textbf{Voxels } & \textbf{Axons }           \\ 
    \hline\hline
    
    \multirow{2}{*}{TBI1} & Contralateral & $14 \times 14 \times 15$ &$13.75 \times 13.75 \times 50$ & $1024 \times 1024 \times 300$ & 41  \\
     & Ipsilateral & $19.1 \times 19.3 \times 15$ & $18.33 \times 18.33 \times 50$ & $1044 \times 1054 \times 300$  & 52 \\ 
    \hline
    \multirow{2}{*}{TBI2} & Contralateral & $14 \times 14 \times 15$ & $13.75 \times 13.75 \times 50$ & $1024 \times 1024 \times 300$ & 67  \\
     & Ipsilateral & $16.9 \times 16.9 \times 15$ & $16.5 \times 16.5 \times 50$ & $1024 \times 1024 \times 300$ & 18  \\ 
    \hline
    \multirow{2}{*}{Sham1} & Contralateral & $17.2 \times 17.2 \times 15$ & $16.81 \times 16.81 \times 50$  & $1024 \times 1024 \times 300$ & 133 \\
     & Ipsilateral & $14 \times 14 \times 15$ & $13.75 \times 13.75 \times 50$ & $1024 \times 1024 \times 300$ & 57 \\ \hline
    \end{tabular}
    \label{dataset_table}    
\end{table*}

\section{Results} \label{section_results}
In this section, we use gACSON to study axonal morphometry in gray matter using a novel dataset consisting of six 3D SBEM volumes. Tissue samples were obtained from the somatosensory cortex of one sham-operated and two TBI rats, both ipsi- and contralateral to the sham operation or injury. Our aim is to investigate the persistent morphological changes of myelinated axons after brain injury.

\subsection{Datasets} \label{section_materials}
\revised{We used three adult male Sprague-Dawley rats (10-weeks old, weight of \SI{320}-\SI{380}{\gram}, Harlan Netherlands B.V., Horst, Netherlands). The rats were accommodated in separated cages in an environment of \SI{22}{\celsius} temperature, 60$\%$ humidity, 12 h of light per day, and unlimited water and food. Animal Care and Use Committee of the Provincial Government of Southern Finland authorized all the actions performed on the animals accordingly to the ground rules addressed by the European Community Council Directive 86/609/EEC. 

TBI was induced by lateral fluid percussion injury in two rats, as described in \cite{Kharatishvili2006} using a fluid-percussion device. The impact pressure was adjusted to a 3.2-3.4 atm to induce a severe injury. The sham-operated rat underwent all the surgical procedures except the impact. After five months, the rats were transcardially perfused using 0.9$\%$ NaCl followed by 4$\%$ paraformaldehyde and the brains were post-fixed in 4$\%$ paraformaldehyde 1$\%$ glutaraldehyde overnight at \SI{4}{\celsius}. Then, the brains were sectioned into 1-mm thick coronal sections and a section at approximately -3.80 mm from bregma from each brain were further dissected into smaller samples containing the areas of interest. \revised{We collected two samples from each brain: the ipsilateral and contralateral hemispheres of the layer VI of the primary somatosensory cortex.} The samples were stained using an enhanced staining protocol described in \cite{Deerinck2010}.

After selecting the area of interest within the samples, the blocks were further trimmed and imaged in a scanning electron microscope (Quanta 250 Field Emission Gun; FEI Co., Hillsboro, OR, USA), equipped with the 3View system (Gatan Inc., Pleasanton, CA, USA) using a backscattered electron detector (Gatan Inc.). The face of the blocks was the \textit{x-y} plane, and the \textit{z}-axis was the cutting direction. All blocks were imaged using a beam voltage of \SI{2.5}{\kilo\volt} and a pressure of 0.15 Torr. Microscopy Image Browser \cite{Belevich2016} (MIB; \url{http://mib.helsinki.fi}) was used to find the bounding box of the collected 2D-EM images and run the voxel-size calibration. The collected images were further aligned by measuring the translation between the consecutive images using the cross-correlation cost function (MIB, \textit{drift correction}). Finally, the contrast of the collected images was normalized such that the mean and standard deviation of the histogram of each image matches the mean and standard deviation of the whole image stack. Information about the 3D-EM data are shown in Table \ref{dataset_table}.}

\begin{figure}[t]
\centering
\includegraphics[width=1\columnwidth]{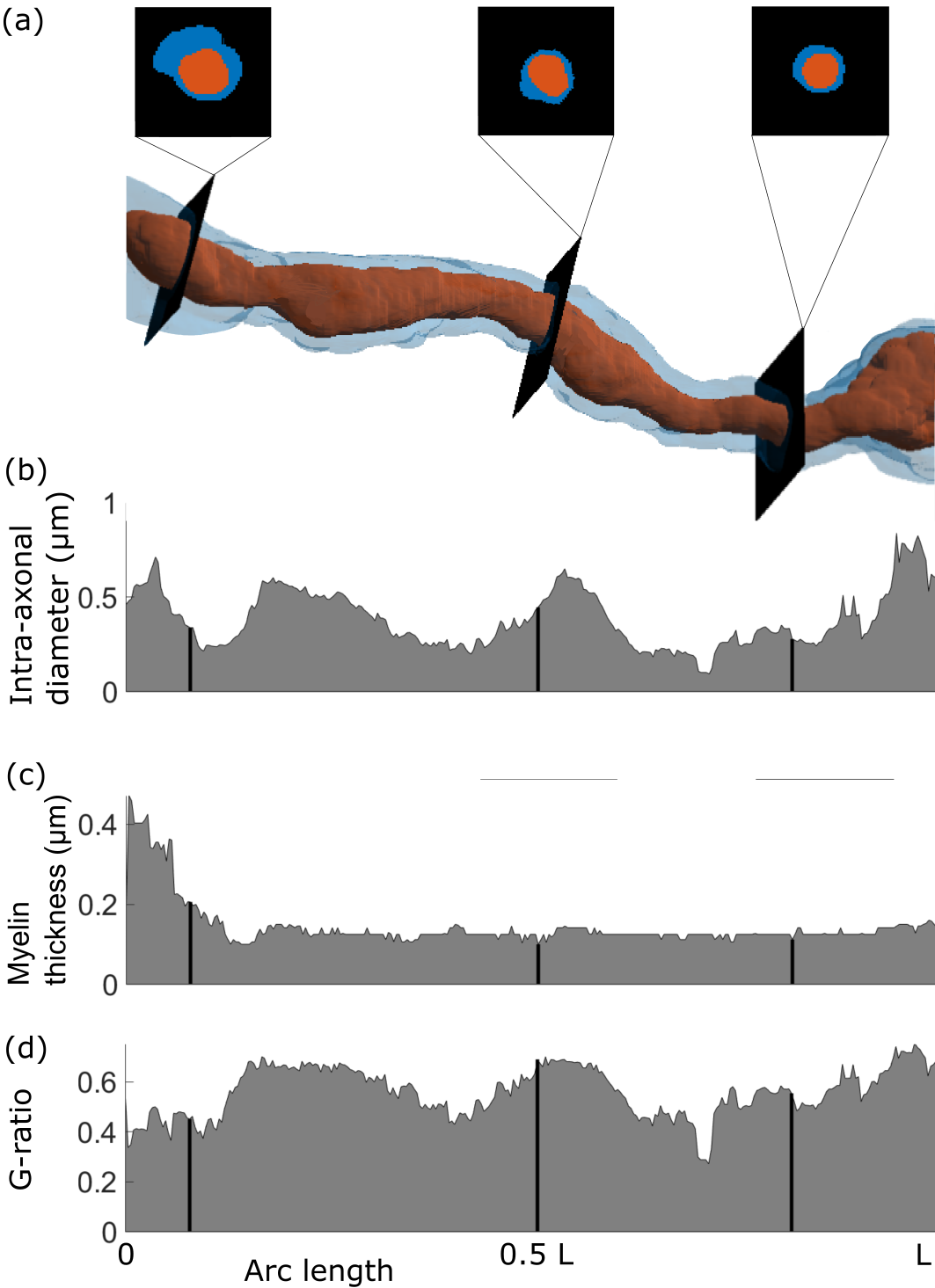}
\caption{Morphology analysis of a myelinated axon. (\textbf{a}) A myelinated axon with cross-sections perpendicular to its skeleton; myelin is blue and the intra-axonal space is red. (\textbf{b}) The diameter of the intra-axonal space \revised{along the length of the axon (L)}: mean = $\SI{0.37}{\micro\meter}$ and std = $\SI{0.15}{\micro\meter}$. (\textbf{c}) The myelin thickness along the axonal skeleton: mean = $\SI{0.15}{\micro\meter}$ and std = $\SI{0.06}{\micro\meter}$. (\textbf{d}) The g-ratio along the axonal skeleton: mean = $0.56$ and std = $0.1$.}
\label{cs_analysis}
\end{figure}

\begin{table*}[t]
    \centering
    \caption{Comparison of myelinated axon morphology between sham-operated and TBI rats. The mean and 95$\%$ confidence interval of each group are denoted as $\mu$ and 95$\%$ CI, respectively. Eq diameter and myl thickness represent equivalent diameter and myelin thickness, respectively. Confidence intervals were calculated with the bias-corrected and accelerated percentile bootstrap method using the Matlab function bootci. The tstat column indicates the Welch t-statistic value for the group difference. The asterisk indicates a statistically significant difference between groups. \revised{The reported p-values have not been corrected for multiple comparisons.} }
    
    \label{tbl:stats}
    \begin{tabular}{l|lllllll}
    {\cellcolor[rgb]{0.937,0.937,0.937}}\textbf{Hemisphere} & 
    {\cellcolor[rgb]{0.937,0.937,0.937}}\begin{tabular}[c]{@{}l@{}} \textbf{Morphology}\\ \textbf{parameter}\end{tabular} & {\cellcolor[rgb]{0.937,0.937,0.937}}\textbf{$\mu_{\text{Sham}}$} & {\cellcolor[rgb]{0.937,0.937,0.937}}\textbf{95$\%$ CI \textsubscript{Sham}} & {\cellcolor[rgb]{0.937,0.937,0.937}}\textbf{$\mu_{\text{TBI}}$} & {\cellcolor[rgb]{0.937,0.937,0.937}}\textbf{95$\%$ CI \textsubscript{TBI}} &    
    {\cellcolor[rgb]{0.937,0.937,0.937}}\textbf{tstat} & 
    {\cellcolor[rgb]{0.937,0.937,0.937}}\textbf{\textit{p}-value}\\ \hline\hline
    
    \multirow{6}{*}{Ipsilateral} 
    & Eq diameter   & 0.48 & [0.44, 0.54] & 0.42 & [0.39, 0.46] & 2.12 & 0.032*\\ \cline{2-8}
    & Minor axis    & 0.39 & [0.36, 0.43] &	0.35 & [0.33, 0.38] & 1.68 & 0.088 \\ \cline{2-8}
    & Major axis    & 0.65 & [0.58, 0.75] &	0.56 & [0.51, 0.65] & 1.71 & 0.082 \\ \cline{2-8}
    & Eccentricity  & 0.74 & [0.71, 0.77] & 0.74 & [0.72, 0.75] & 0.17 & 0.854 \\ \cline{2-8}
    & Myl thickness & 0.15 & [0.14, 0.16] & 0.14 & [0.13, 0.15] & 1.18 & 0.250 \\ \cline{2-8}
    & G-ratio       & 0.61 & [0.59, 0.63] &	0.61 & [0.58, 0.65] & 0.00 & 0.999 \\ \hline\hline
    
    \multirow{6}{*}{Contralateral} 
    & Eq diameter   & 0.59 & [0.55, 0.65] &	0.55 & [0.51, 0.61] & 1.08 & 0.280 \\ \cline{2-8}
    & Minor axis    & 0.48 & [0.44, 0.52] &	0.44 & [0.40, 0.48] & 1.50 & 0.143 \\ \cline{2-8}
    & Major axis    & 0.77 & [0.71, 0.85] &	0.75 & [0.68, 0.83] & 0.49 & 0.631 \\ \cline{2-8}
    & Eccentricity  & 0.75 & [0.74, 0.77] &	0.77 & [0.76, 0.79] & -1.87& 0.065 \\ \cline{2-8}
    & Myl thickness & 0.16 & [0.15, 0.17] &	0.16 & [0.15, 0.17] & 0.78 & 0.436 \\ \cline{2-8}
    & G-ratio       & 0.63 & [0.62, 0.65] &	0.63 & [0.61, 0.65] & 0.20 & 0.843 \\ \hline
    
    \end{tabular}
\end{table*}

\begin{table}[t]
    \centering
    \caption{Comparison of the morphology of myelinated axons between contralateral and ipsilateral hemispheres in sham-operated and in TBI rats. See Table  \ref{tbl:stats} for notation.}
    
    \label{tbl:stats_contra_ipsi}
    \begin{tabular}{l|lll}
    {\cellcolor[rgb]{0.937,0.937,0.937}}\textbf{Group} & 
    {\cellcolor[rgb]{0.937,0.937,0.937}}\begin{tabular}[c]{@{}l@{}} \textbf{Morphology}\\ \textbf{parameter}\end{tabular} & {\cellcolor[rgb]{0.937,0.937,0.937}}\textbf{tstat} & {\cellcolor[rgb]{0.937,0.937,0.937}}\textbf{\textit{p}-value}\\ \hline\hline
    
    \multirow{6}{*}{Sham} 
    & Eq diameter   & 3.17 & 0.005*\\ \cline{2-4}
    & Minor axis    & 3.38 & 0.003*\\ \cline{2-4}
    & Major axis    & 2.16 & 0.056\\ \cline{2-4}
    & Eccentricity  & 0.82 & 0.414 \\ \cline{2-4}
    & Myl thickness & 2.41 & 0.021*\\ \cline{2-4}
    & G-ratio       & 1.86 & 0.065 \\ \hline\hline
    
    \multirow{6}{*}{TBI} 
    & Eq diameter   & 4.43 & 0.001*\\ \cline{2-4}
    & Minor axis    & 3.59 & 0.001*\\ \cline{2-4}
    & Major axis    & 3.71 & 0.001*\\ \cline{2-4}
    & Eccentricity  & 3.00 & 0.002*\\ \cline{2-4}
    & Myl thickness & 2.35 & 0.018*\\ \cline{2-4}
    & G-ratio        & 1.08 & 0.294 \\ \hline    
    \end{tabular}
\end{table}

\subsection{Segmentation pipeline}
We used BM4D to denoise the six pre-processed 3D SBEM data of the primary somatosensory cortex (section \ref{bm4d}). For the semantic segmentation of myelin, we applied the bounded volume growing algorithm, where the initial seeds are generated randomly (section \ref{myelin_semantic}). We segmented the intra-axonal space of myelinated axons using the bounded volume growing algorithm, where the initial seeds were generated using the watershed centroids method (section \ref{axon_instance}). We set the $H$ for the H-maxima transform equal to two, the similarity threshold of the volume growing method to 0.1, and excluded those connected components with an area smaller than 100 voxels and larger than one million voxels. After the semantic segmentation of myelin and the instance segmentation of the intra-axonal spaces of myelinated axons, we performed the instance segmentation of myelin (section \ref{myelin_instance}) to assign myelin to its corresponding intra-axonal space. Figure \ref{seg_volume} shows the 3D rendering of the intra-axonal space of myelinated axons from a TBI dataset. From the six SBEM volumes, the pipeline segmented 368 axons that were longer than \SI{4}{\micro\meter} as shown in Table 2.

\subsection{Morphometry pipeline}
We analyzed the morphology of the segmented myelinated axons using our morphometry algorithm described in section \ref{morph_pip}. We extracted the minor and major diameters of fitted ellipses, equivalent diameter, eccentricity, myelin thickness, and g-ratio for every cross-section along all myelinated axons in all 3D-EM data. Figure \ref{cs_analysis}a shows a myelinated axon with three cross-sections perpendicular to its axonal skeleton. The cross-sectional diameter of the intra-axonal space of the myelinated axon is shown along its length, where the mean is equal to \SI{0.37}{\micro\meter}, and the standard deviation is equal to \SI{0.15}{\micro\meter} (Figure \ref{cs_analysis}b), indicating considerable variation of axonal diameter along the axon. Figure \ref{cs_analysis}c shows the myelin thickness along the length of the myelinated axon where the mean value equals \SI{0.15}{\micro\meter} and the standard deviation equals \SI{0.06}{\micro\meter}. The g-ratio along the length of the myelinated axon (Figure \ref{cs_analysis}d), where the mean value is equal to 0.56, and the standard deviation is equal to 0.1.

\subsection{Morphology of myelinated axons in the somatosensory cortex of sham-operated and TBI rats} \label{Statistics}

We quantified the morphology of myelinated axons in the somatosensory cortex to investigate the persistent changes in myelinated axon morphology after TBI. To do this, we first established a threshold for the length of myelinated axons and selected for analysis those myelinated axons that were long enough to traverse approximately one-third of the SBEM volume (longer than \SI {4}{\micro\meter}). Thresholding myelinated axons based on their length excluded those axons that were incorrectly split or components that were incorrectly labelled as myelinated axons. \revised{It would be possible to exclude axons also manually, but we preferred to set a length-threshold to increase the objectivity and repeatability of the analysis.}  We represented an axonal parameter by the median of its cross-sectional values. 

To compare the morphology of myelinated axons in sham-operated and TBI animals, we performed the two-sample Studentized permutation test using the Welch t-statistic as the test statistic \cite{janssen1997studentized}.The test was performed separately for both hemispheres. Because we had only one animal in the sham-operated group, we pooled the axonal measurements from the two animals in the TBI group. The alpha threshold for statistical significance was set at 0.05 for all analyses. The analysis was performed using custom code in Matlab R2018b. Similarly, we assessed differences between axons in the ipsi and contralateral hemispheres.\revised{We provide here uncorrected p-values for clarity as 1) our focus on quantification, not hypothesis testing, 2) there are several alternatives to set the number of tests (i.e., should we correct for tests between Sham and TBI and the tests between the hemispheres or should we consider difference between the Sham and TBI and hemispheric differences as separate experiments), and 3) knowing that p-values are uncorrected, the readers can adjust their expectations \cite{gelman2012we}}. 

The results of the analysis are shown in \revised{Supplementary} Figure 1 and Table \ref{tbl:stats}. For example, the average equivalent diameter of ipsilateral axons was \SI{0.48}{\micro\meter} (95\% CI [\SI{0.43}{\micro\meter} \SI{0.53}{\micro\meter}] in the sham-operated animal and \SI{0.48}{\micro\meter} (95\% CI [\SI{0.38}{\micro\meter} \SI{0.45}{\micro\meter}] in the injured animals. 
Our analysis shows that the equivalent diameter of the intra-axonal space of myelinated axons was significantly smaller in the TBI animals when the groups were compared ipsilaterally (Table \ref{tbl:stats}, \revised{Supplementary} Figure 1) \revised{although if any multiple comparisons correction took place that effect would disappear}. We did not find the same effect on equivalent diameter contralaterally (Table \ref{tbl:stats}, \revised{Supplementary} Figure 1). We found no significant difference between minor and major axis length, eccentricity, myelin thickness, or g-ratio when comparing sham-operated and TBI rats ipsi- and contralaterally (Table \ref{tbl:stats}, \revised{Supplementary} Figure 1). The comparison between ipsi- and contralateral hemispheres is shown in Table \ref{tbl:stats_contra_ipsi}. Equivalent diameter, minor axis, and myelin thickness were greater in the contralateral hemisphere in both sham-operated and TBI rats. In addition, eccentricity and major axis were greater in the contralateral hemisphere of animals after TBI. \revised{Especially in TBI rats, the hemisphere effects were strong and would survive multiple comparison correction.} 

\subsection{Computation time and memory consumption}
Table \ref{comp_time} shows the computation time of the image segmentation and morphometry for the SBEM images used in this study. Using a Hp EliteDesk 800 G4 TWRWS (Windows 10 (64-bit) operating system, an Intel(R) Core(TM) i7-8700K CPU @3.70 GHz processor, and 64 GB RAM), gACSON averaged 5.7 hours for segmentation and 2.4 hours for morphometry for our SBEM datasets. The differences in computation time across the datasets are due to differences in the number of seeds evaluated for BVG during the segmentation and differences in the number and lengths of axons quantified during the morphometry. \revised{The BVG algorithm is written in “plain” Matlab-code optimized for speed, while BM4D (\url{https://webpages.tuni.fi/foi/GCF-BM3D}), SLIC (\url{https://github.com/fk128/SLICSupervoxels}), and multi-stencils fast marching and Runge-Kutta method used in the skeletonization process (\url{https://github.com/cbm755/fast_marching_kroon}) rely on compiled MEX files. Note that to reduce the computation time advantages of compiled programming languages such as C and C++ compared to interpreted programming languages such as Matlab, the Matlab execution engine uses just-in-time compilation of all Matlab code with a single execution pathway that considerably speeds-up the 3D voxel-based processing.} \revised{The peak of the memory consumption of the pipeline occurred after calculating SLIC supervoxels when the pipeline consumed 11 GB of memory from which approximately 6.7 GB was consumed by SLIC and the rest was due to maintaining multiple copies of data in the memory.}

\begin{table}[t]
    \centering
    \caption{The computation time of gACSON segmentation (Seg), seed location are defined using the watershed centroids, and the gACSON morphology (Morph) analysis.}    
    \label{comp_time}
    \begin{tabular}{l|lcc}
    {\cellcolor[rgb]{0.937,0.937,0.937}}\textbf{Rat} & 
    {\cellcolor[rgb]{0.937,0.937,0.937}}\textbf{Hemisphere } & {\cellcolor[rgb]{0.937,0.937,0.937}}\textbf{Seg (h)} & {\cellcolor[rgb]{0.937,0.937,0.937}}\textbf{Morph (h)}\\ \hline\hline
    
    \multirow{2}{*}{TBI1} & Contralateral & 5 & 1.5 \\
                          & Ipsilateral & 3 & 1.6   \\ \hline
    \multirow{2}{*}{TBI2} & Contralateral & 4 & 3.3 \\
                          & Ipsilateral & 6 & 0.8   \\ \hline
    \multirow{2}{*}{Sham1}& Contralateral & 8 & 4.4 \\
                          & Ipsilateral & 8 & 2.7   \\ \hline
    \end{tabular}
\end{table}

\section{Discussion} \label{section_Discussion}

In this paper, we present a GUI-based gACSON software for automatic visualization, segmentation and analysis of myelinated axon morphology in 3D- EM data. We developed \revised{Matlab-based} gACSON to provide an extensible open-source platform for both automatic and semi-automatic segmentation of brain tissue ultrastructure in 3D- EM data. The source code is available under the MIT license at \url{https://github.com/AndreaBehan/g-ACSON}. To demonstrate the applicability of the \revised{software}, we performed instance segmentation of myelin and the intra-axonal space of myelinated axons in six SBEM datasets from the somatosensory cortex of one sham-operated and two TBI rats.

gACSON provides a framework for automatic segmentation of brain tissue ultrastructures and a means for manual segmentation and correction of automatic segmentations. The segmentation pipeline is easy to use even for inexperienced users. Only a few intuitive parameters need to be set, such as the similarity threshold or the maximum and minimum volume of a segmented component. \revised{The software} provides several methods for semantic segmentation of myelin that are applicable in different situations. \revised{Importantly, it} enables instance segmentation of myelinated axons and provides several methods for generating the initial seeds for the BVG algorithm \cite{Abdollahzadeh2019AutomatedMatter}, which extend our previous algorithms for axon segmentation \cite{Abdollahzadeh2019AutomatedMatter,Abdollahzadeh2021DeepACSONMicroscopy}.

Axon instance segmentation makes no assumptions about the shape of an axon, unlike axon and myelin segmentation tools such as AxonSeg \cite{Zaimi2016AxonSeg:Analysis} or MyelTracer \cite{Kaiser2021Myeltracer:Quantification}. AxonSeg and MyelTracer work on 2D- EM data with a tissue section perpendicular to the main axonal orientation, so that the cross sections of the axons are approximately circular. Thus, these tools assume that axons are highly parallel and that there is a main axonal orientation against which the tissue is cut perpendicularly. However, this assumption is not generally true: axons can orient, cross, or bend in any direction \cite{Abdollahzadeh2021DeepACSONMicroscopy}. Thus, while some axons can be cut perpendicular to their skeleton, others can be cut obliquely and along their length. Unlike MyelTracer \cite{Kaiser2021Myeltracer:Quantification}, which separates adjacent myelin sheaths based on user input, gACSON performs this task automatically. The gACSON segmentation pipeline enables automatic instance segmentation of myelin, linking the intra-axonal space to its surrounding myelin sheath. This is a unique feature compared to existing software tools such as \cite{Belevich2016, Cardona2012, Knossos, Saalfeld2009, Berning2015, ilastik}.

Several functionalities of gACSON, such as browsing 3D data, proofreading segmentations, or evaluating and correcting a segmentation result, are not limited to EM data, but can be used for any 2D or 3D image data. However, the segmentation and morphometry pipelines are specifically designed for the analysis of myelinated axons in 3D-EM of gray or white matter. \revised{They} extract the morphology of the segmented myelinated axons in 3D, allowing accurate analysis of the shape of the myelinated axons. These measurements are important for studying the conduction velocity of myelinated axons in electrophysiology \cite{gRatio} or for studying the characterization of axonal diameter with magnetic resonance imaging \cite{Salo2018QuantificationBrain}. We illustrated the use of gACSON by analyzing an unpublished dataset of 3D SBEM volumes from rat somatosensory cortex.

Our morphology analyses showed that the equivalent diameter of ipsilateral axons was significantly smaller in TBI rats than in sham-operated rats. Obviously, this result should be interpreted with caution because of the small number of animals available. However, the result is consistent with the decreased diameter of myelinated axons in the same animals in both the corpus callosum and cingulum \cite{Abdollahzadeh2021DeepACSONMicroscopy}. \revised{Also, despite the volumes included in this study were selected in a specific area, the layer VI of the primary somatosensory cortex, variations in the cellular content of these small volumes may occur, which might affect the comparative results.}

\revised{
Certain limitations and challenges need to be addressed in future software updates. For example, the size of datasets cannot exceed the RAM size because the segmentation pipeline loads the whole dataset into memory. For a very large EM volume this requires tiling EM datasets into sub-volumes and then performing a segmentation task. This task should be automated. The instance segmentation of myelin may include inaccuracies when adjacent myelin sheaths have a substantial difference in thickness as the current watershed-based segmentation divides the myelin content equally between two adjacent axons. We plan to use distance transform with a propagation speed relative to myelin thickness to remedy such scenarios.}

\section*{Acknowledgments}
We would like to thank Maarit Pulkkinen for her assistance with the animal handling, Mervi Lindman and Antti Salminen for their help preparing the samples for SBEM. 
 
\section*{Funding}
This research has been funded in part by the Academy of Finland grant $\#$316258 (JT) and $\#$323385 (AS), European Social Fund project S21770 (JT), and Biocenter Finland and University of Helsinki (IB, EJ, and SBEM imaging).

\section*{Competing interest statement}
The authors declare that they have no competing interests.

\section*{Availability of data and material}
The source code of gACSON is available at \url{https://github.com/AndreaBehan/g-ACSON} under the MIT License.  The code is also archived at Zenodo DOI: 10.5281/zenodo.3693563.  Quantitative axon morphology measures are attached as a supplementary file to this manuscript. The 3D-EM data and their segmentations will be made available at \url{https://doi.org/10.23729/bad417ca-553f-4fa6-ae0a-22eddd29a230}.  

\bibliography{mybibfile}

\begin{thebibliography}{10}
\expandafter\ifx\csname url\endcsname\relax
  \def\url#1{\texttt{#1}}\fi
\expandafter\ifx\csname urlprefix\endcsname\relax\def\urlprefix{URL }\fi
\expandafter\ifx\csname href\endcsname\relax
  \def\href#1#2{#2} \def\path#1{#1}\fi

\bibitem{Phelps2021ReconstructionMicroscopy}
J.~S. Phelps, D.~G.~C. Hildebrand, B.~J. Graham, A.~T. Kuan, L.~A. Thomas,
  T.~M. Nguyen, J.~Buhmann, A.~W. Azevedo, A.~Sustar, S.~Agrawal, M.~Liu, B.~L.
  Shanny, J.~Funke, J.~C. Tuthill, W.~C.~A. Lee,
  \href{https://doi.org/10.1016/j.cell.2020.12.013}{{Reconstruction of motor
  control circuits in adult Drosophila using automated transmission electron
  microscopy}}, Cell 184~(3) (2021) 759--774.
\newblock \href {https://doi.org/10.1016/j.cell.2020.12.013}
  {\path{doi:10.1016/j.cell.2020.12.013}}.
\newline\urlprefix\url{https://doi.org/10.1016/j.cell.2020.12.013}

\bibitem{Nahirney2021}
P.~C. Nahirney, M.-E. Tremblay,
  \href{https://www.frontiersin.org/article/10.3389/fcell.2021.629503}{Brain
  ultrastructure: Putting the pieces together}, Frontiers in Cell and
  Developmental Biology 9 (2021) 187.
\newblock \href {https://doi.org/10.3389/fcell.2021.629503}
  {\path{doi:10.3389/fcell.2021.629503}}.
\newline\urlprefix\url{https://www.frontiersin.org/article/10.3389/fcell.2021.629503}

\bibitem{Abdollahzadeh2021DeepACSONMicroscopy}
A.~Abdollahzadeh, I.~Belevich, E.~Jokitalo, A.~Sierra, J.~Tohka,
  \href{http://dx.doi.org/10.1038/s42003-021-01699-w
  http://www.nature.com/articles/s42003-021-01699-w}{{DeepACSON automated
  segmentation of white matter in 3D electron microscopy}}, Communications
  Biology 4~(1) (2021) 179.
\newblock \href {https://doi.org/10.1038/s42003-021-01699-w}
  {\path{doi:10.1038/s42003-021-01699-w}}.
\newline\urlprefix\url{http://dx.doi.org/10.1038/s42003-021-01699-w
  http://www.nature.com/articles/s42003-021-01699-w}

\bibitem{Kleinnijenhuis2020AMicroscopy}
M.~Kleinnijenhuis, E.~Johnson, J.~Mollink, S.~Jbabdi, K.~L. Miller, {A
  semi-automated approach to dense segmentation of 3D white matter electron
  microscopy}, bioRxiv (2020).
\newblock \href {https://doi.org/10.1101/2020.03.19.979393}
  {\path{doi:10.1101/2020.03.19.979393}}.

\bibitem{Yuan2021HIVE-Net:Images}
Z.~Yuan, X.~Ma, J.~Yi, Z.~Luo, J.~Peng,
  \href{https://doi.org/10.1016/j.cmpb.2020.105925}{{HIVE-Net: Centerline-aware
  hierarchical view-ensemble convolutional network for mitochondria
  segmentation in EM images}}, Computer Methods and Programs in Biomedicine 200
  (2021) 105925.
\newblock \href {https://doi.org/10.1016/j.cmpb.2020.105925}
  {\path{doi:10.1016/j.cmpb.2020.105925}}.
\newline\urlprefix\url{https://doi.org/10.1016/j.cmpb.2020.105925}

\bibitem{Berning2015}
M.~Berning, K.~Boergens, M.~Helmstaedter,
  \href{https://doi.org/10.1016/j.neuron.2015.09.003}{{SegEM}: Efficient image
  analysis for high-resolution connectomics}, Neuron 87~(6) (2015) 1193--1206.
\newblock \href {https://doi.org/10.1016/j.neuron.2015.09.003}
  {\path{doi:10.1016/j.neuron.2015.09.003}}.
\newline\urlprefix\url{https://doi.org/10.1016/j.neuron.2015.09.003}

\bibitem{Abdollahzadeh2019AutomatedMatter}
A.~Abdollahzadeh, I.~Belevich, E.~Jokitalo, J.~Tohka, A.~Sierra, {Automated 3D
  Axonal Morphometry of White Matter}, Scientific Reports 9~(1) (2019) 6084.
\newblock \href {https://doi.org/10.1038/s41598-019-42648-2}
  {\path{doi:10.1038/s41598-019-42648-2}}.

\bibitem{Belevich2016}
I.~Belevich, M.~Joensuu, D.~Kumar, H.~Vihinen, E.~Jokitalo,
  \href{https://doi.org/10.1371/journal.pbio.1002340}{Microscopy image browser:
  A platform for segmentation and analysis of multidimensional datasets},
  {PLOS} Biology 14~(1) (2016) e1002340.
\newblock \href {https://doi.org/10.1371/journal.pbio.1002340}
  {\path{doi:10.1371/journal.pbio.1002340}}.
\newline\urlprefix\url{https://doi.org/10.1371/journal.pbio.1002340}

\bibitem{Belevich2021DeepMIB:Segmentation}
I.~Belevich, E.~Jokitalo,
  \href{https://journals.plos.org/ploscompbiol/article?id=10.1371/journal.pcbi.1008374}{{DeepMIB:
  User-friendly and open-source software for training of deep learning network
  for biological image segmentation}}, PLOS Computational Biology 17~(3) (2021)
  e1008374.
\newblock \href {https://doi.org/10.1371/JOURNAL.PCBI.1008374}
  {\path{doi:10.1371/JOURNAL.PCBI.1008374}}.
\newline\urlprefix\url{https://journals.plos.org/ploscompbiol/article?id=10.1371/journal.pcbi.1008374}

\bibitem{Knossos}
M.~Helmstaedter, K.~Briggman, W.~Denk, High-accuracy neurite reconstruction for
  high-throughput neuroanatomy., Nature Neuroscience (2011) 1081–1088\href
  {https://doi.org/10.1038/nn.2868} {\path{doi:10.1038/nn.2868}}.

\bibitem{Boergens2017WebKnossos:Connectomics}
K.~M. Boergens, M.~Berning, T.~Bocklisch, D.~Br{\"{a}}unlein, F.~Drawitsch,
  J.~Frohnhofen, T.~Herold, P.~Otto, N.~Rzepka, T.~Werkmeister, D.~Werner,
  G.~Wiese, H.~Wissler, M.~Helmstaedter,
  \href{http://dx.doi.org/10.1038/nmeth.4331}{{WebKnossos: Efficient online 3D
  data annotation for connectomics}}, Nature Methods 14~(7) (2017) 691--694.
\newblock \href {https://doi.org/10.1038/nmeth.4331}
  {\path{doi:10.1038/nmeth.4331}}.
\newline\urlprefix\url{http://dx.doi.org/10.1038/nmeth.4331}

\bibitem{Zaimi2016AxonSeg:Analysis}
A.~Zaimi, T.~Duval, A.~Gasecka, D.~C{\^{o}}t{\'{e}}, N.~Stikov, J.~Cohen-Adad,
  {AxonSeg: Open source software for axon and myelin segmentation and
  morphometric analysis}, Frontiers in Neuroinformatics 10~(AUG) (2016) 1--13.
\newblock \href {https://doi.org/10.3389/fninf.2016.00037}
  {\path{doi:10.3389/fninf.2016.00037}}.

\bibitem{Zaimi2018AxonDeepSeg:Networks}
A.~Zaimi, M.~Wabartha, V.~Herman, P.-L. Antonsanti, C.~S. Perone,
  J.~Cohen-Adad,
  \href{http://www.nature.com/articles/s41598-018-22181-4}{{AxonDeepSeg:
  automatic axon and myelin segmentation from microscopy data using
  convolutional neural networks}}, Scientific Reports 8~(1) (2018) 3816.
\newblock \href {https://doi.org/10.1038/s41598-018-22181-4}
  {\path{doi:10.1038/s41598-018-22181-4}}.
\newline\urlprefix\url{http://www.nature.com/articles/s41598-018-22181-4}

\bibitem{Cardona2012}
A.~Cardona, S.~Saalfeld, J.~Schindelin, I.~Arganda-Carreras, S.~Preibisch,
  M.~Longair, P.~Tomancak, V.~Hartenstein, R.~Douglas,
  \href{https://doi.org/10.1371/journal.pone.0038011}{{TrakEM}2 software for
  neural circuit reconstruction}, {PLoS} {ONE} 7~(6) (2012) e38011.
\newblock \href {https://doi.org/10.1371/journal.pone.0038011}
  {\path{doi:10.1371/journal.pone.0038011}}.
\newline\urlprefix\url{https://doi.org/10.1371/journal.pone.0038011}

\bibitem{Saalfeld2009}
S.~Saalfeld, A.~Cardona, V.~Hartenstein, P.~Tomancak,
  \href{https://doi.org/10.1093/bioinformatics/btp266}{{CATMAID}: collaborative
  annotation toolkit for massive amounts of image data}, Bioinformatics 25~(15)
  (2009) 1984--1986.
\newblock \href {https://doi.org/10.1093/bioinformatics/btp266}
  {\path{doi:10.1093/bioinformatics/btp266}}.
\newline\urlprefix\url{https://doi.org/10.1093/bioinformatics/btp266}

\bibitem{Berger2018VASTStacks}
D.~R. Berger, H.~S. Seung, J.~W. Lichtman, {VAST (Volume Annotation and
  Segmentation Tool): Efficient manual and semi-automatic labeling of large 3D
  image stacks}, Frontiers in Neural Circuits 12~(October) (2018).
\newblock \href {https://doi.org/10.3389/fncir.2018.00088}
  {\path{doi:10.3389/fncir.2018.00088}}.

\bibitem{Jorstad2018NeuroMorph:Connectivity}
A.~Jorstad, J.~Blanc, G.~Knott, {NeuroMorph: A Software Toolset for 3D Analysis
  of Neurite Morphology and Connectivity}, Frontiers in Neuroanatomy 12~(July)
  (2018) 1--12.
\newblock \href {https://doi.org/10.3389/fnana.2018.00059}
  {\path{doi:10.3389/fnana.2018.00059}}.

\bibitem{ilastik}
C.~Sommer, C.~Straehle, U.~Kothe, F.~Hamprecht, ilastik: Interactive learning
  and segmentation toolkit, in: Eighth IEEE International Symposium on
  Biomedical Imaging (ISBI 2011).Proceedings, 2011, pp. 230--233.
\newblock \href {https://doi.org/10.1109/ISBI.2011.5872394}
  {\path{doi:10.1109/ISBI.2011.5872394}}.

\bibitem{Kaiser2021Myeltracer:Quantification}
T.~Kaiser, H.~M. Allen, O.~Kwon, B.~Barak, J.~Wang, Z.~He, M.~Jiang, G.~Feng,
  {Myeltracer: A semi-automated software for myelin g-ratio quantification},
  eNeuro 8~(4) (2021) 1--9.
\newblock \href {https://doi.org/10.1523/ENEURO.0558-20.2021}
  {\path{doi:10.1523/ENEURO.0558-20.2021}}.

\bibitem{bm4D}
M.~Maggioni, V.~Katkovnik, K.~Egiazarian, A.~Foi, Nonlocal transform-domain
  filter for volumetric data denoising and reconstruction, IEEE Transactions on
  Image Processing 22~(1) (2013) 119--133.
\newblock \href {https://doi.org/10.1109/TIP.2012.2210725}
  {\path{doi:10.1109/TIP.2012.2210725}}.

\bibitem{Sim2006a}
K.~S. Sim, J.~T.~L. Thong, J.~C.~H. Phang, {Effect of shot noise and secondary
  emission noise in scanning electron microscope images}, Scanning 26~(1)
  (2006) 36--40.
\newblock \href {https://doi.org/10.1002/sca.4950260106}
  {\path{doi:10.1002/sca.4950260106}}.

\bibitem{Breiman2001}
L.~Breiman, \href{https://doi.org/10.1023/a:1010933404324}{Random forests},
  Machine Learning 45~(1) (2001) 5--32.
\newblock \href {https://doi.org/10.1023/a:1010933404324}
  {\path{doi:10.1023/a:1010933404324}}.
\newline\urlprefix\url{https://doi.org/10.1023/a:1010933404324}

\bibitem{lee2010image}
S.~H. Lee, H.~I. Koo, N.~I. Cho, Image segmentation algorithms based on the
  machine learning of features, Pattern Recognition Letters 31~(14) (2010)
  2325--2336.

\bibitem{RegionGrowing}
R.~Adams, L.~Bischof, Seeded region growing, IEEE Transactions on Pattern
  Analysis and Machine Intelligence 16~(6) (1994) 641--647.
\newblock \href {https://doi.org/10.1109/34.295913}
  {\path{doi:10.1109/34.295913}}.

\bibitem{Canny1986ADetection}
J.~Canny, {A computational approach to edge detection}, IEEE Transactions on
  Pattern Analysis and Machine Intelligence PAMI-8~(6) (1986) 679--698.
\newblock \href {https://doi.org/10.1109/TPAMI.1986.4767851}
  {\path{doi:10.1109/TPAMI.1986.4767851}}.

\bibitem{Najman1994}
L.~Najman, M.~Schmitt,
  \href{https://doi.org/10.1016/0165-1684(94)90059-0}{Watershed of a continuous
  function}, Signal Processing 38~(1) (1994) 99--112.
\newblock \href {https://doi.org/10.1016/0165-1684(94)90059-0}
  {\path{doi:10.1016/0165-1684(94)90059-0}}.
\newline\urlprefix\url{https://doi.org/10.1016/0165-1684(94)90059-0}

\bibitem{achanta2012slic}
R.~Achanta, A.~Shaji, K.~Smith, A.~Lucchi, P.~Fua, S.~S{\"u}sstrunk, Slic
  superpixels compared to state-of-the-art superpixel methods, IEEE
  transactions on pattern analysis and machine intelligence 34~(11) (2012)
  2274--2282.

\bibitem{Hassouna2005RobustSets}
M.~Hassouna, A.~Farag,
  \href{http://ieeexplore.ieee.org/document/1467303/}{{Robust Centerline
  Extraction Framework Using Level Sets}}, in: 2005 IEEE Computer Society
  Conference on Computer Vision and Pattern Recognition (CVPR'05), Vol.~1,
  IEEE, 2005, pp. 458--465.
\newblock \href {https://doi.org/10.1109/CVPR.2005.306}
  {\path{doi:10.1109/CVPR.2005.306}}.
\newline\urlprefix\url{http://ieeexplore.ieee.org/document/1467303/}

\bibitem{Abdollahzadeh2021CylindricalObjects}
A.~Abdollahzadeh, A.~Sierra, J.~Tohka,
  \href{https://ieeexplore.ieee.org/document/9345673/
  http://dx.doi.org/10.1109/ACCESS.2021.3056958}{{Cylindrical Shape
  Decomposition for 3D Segmentation of Tubular Objects}}, IEEE Access 9 (2021)
  23979--23995.
\newblock \href {https://doi.org/10.1109/ACCESS.2021.3056958}
  {\path{doi:10.1109/ACCESS.2021.3056958}}.
\newline\urlprefix\url{https://ieeexplore.ieee.org/document/9345673/
  http://dx.doi.org/10.1109/ACCESS.2021.3056958}

\bibitem{Kharatishvili2006}
I.~Kharatishvili, J.~Nissinen, T.~McIntosh, A.~Pitk\"{a}nen,
  \href{https://doi.org/10.1016/j.neuroscience.2006.03.012}{A model of
  posttraumatic epilepsy induced by lateral fluid-percussion brain injury in
  rats}, Neuroscience 140~(2) (2006) 685--697.
\newblock \href {https://doi.org/10.1016/j.neuroscience.2006.03.012}
  {\path{doi:10.1016/j.neuroscience.2006.03.012}}.
\newline\urlprefix\url{https://doi.org/10.1016/j.neuroscience.2006.03.012}

\bibitem{Deerinck2010}
T.~Deerinck, E.~Bushong, V.~Lev-Ram, X.~Shu, R.~Tsien, M.~Ellisman,
  \href{https://doi.org/10.1017/s1431927610055170}{Enhancing serial block-face
  scanning electron microscopy to enable high resolution 3-d nanohistology of
  cells and tissues}, Microscopy and Microanalysis 16~(S2) (2010) 1138--1139.
\newblock \href {https://doi.org/10.1017/s1431927610055170}
  {\path{doi:10.1017/s1431927610055170}}.
\newline\urlprefix\url{https://doi.org/10.1017/s1431927610055170}

\bibitem{janssen1997studentized}
A.~Janssen, Studentized permutation tests for non-iid hypotheses and the
  generalized behrens-fisher problem, Statistics \& probability letters 36~(1)
  (1997) 9--21.

\bibitem{gelman2012we}
A.~Gelman, J.~Hill, M.~Yajima, Why we (usually) don't have to worry about
  multiple comparisons, Journal of research on educational effectiveness 5~(2)
  (2012) 189--211.

\bibitem{gRatio}
T.~Chomiak, B.~Hu, \href{https://doi.org/10.1371/journal.pone.0007754}{What is
  the optimal value of the g-ratio for myelinated fibers in the rat {CNS}? {A}
  theoretical approach}, PLOS ONE 4~(11) (2009) 1--7.
\newblock \href {https://doi.org/10.1371/journal.pone.0007754}
  {\path{doi:10.1371/journal.pone.0007754}}.
\newline\urlprefix\url{https://doi.org/10.1371/journal.pone.0007754}

\bibitem{Salo2018QuantificationBrain}
R.~A. Salo, I.~Belevich, E.~Manninen, E.~Jokitalo, O.~Gr{\"{o}}hn, A.~Sierra,
  {Quantification of anisotropy and orientation in 3D electron microscopy and
  diffusion tensor imaging in injured rat brain}, NeuroImage 172~(February)
  (2018) 404--414.
\newblock \href {https://doi.org/10.1016/j.neuroimage.2018.01.087}
  {\path{doi:10.1016/j.neuroimage.2018.01.087}}.

\end{thebibliography}

\renewcommand{\figurename}{Supplementary Figure}
\setcounter{figure}{0}

\begin{figure*}[htp]
\section*{\revised{Supplementary Information}}
\centering
\includegraphics[width=0.92\linewidth]{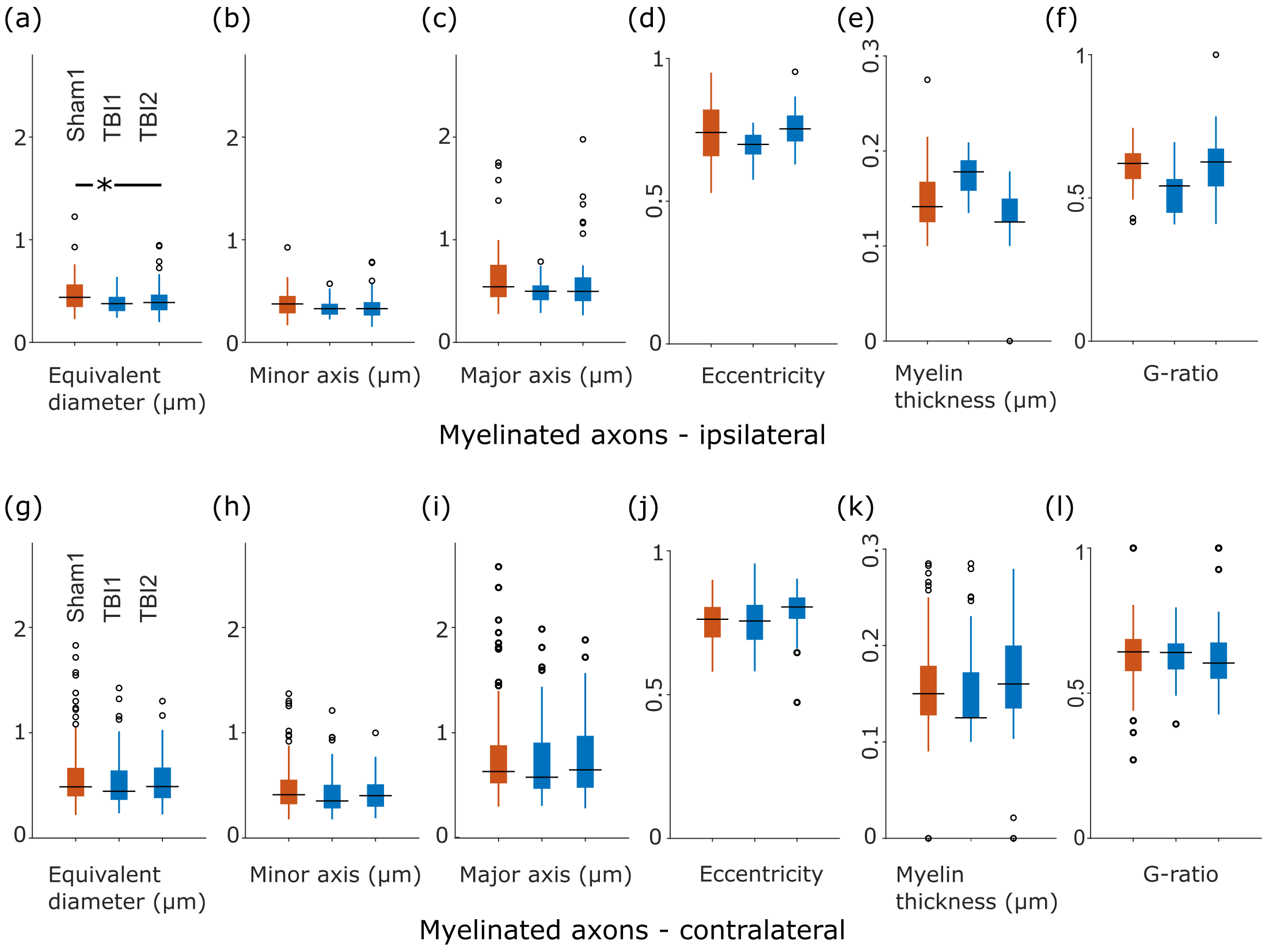}
\caption{The morphology analysis of myelinated axons in the ipsilateral (\textbf{a-f}) and contralateral (\textbf{g-l}) somatosensory cortex of sham-operated and TBI rats: (\textbf{a, g}) equivalent diameter, (\textbf{b, h}) minor axis length (\textbf{c, i}) major axis length, (\textbf{d, j}) axonal eccentricity, (\textbf{e, k}) myelin thickness, and (\textbf{f, l}) g-ratio. On each box, the central mark indicates the median, and the bottom and top edges of the box indicate the $25^{th}$ and $75^{th}$ percentiles, respectively. The whiskers extend to the most extreme data points not considered outliers, and the outliers are plotted individually using the ‘o’ symbol. The asterisk in (\textbf{a}) shows that there is a significant reduction in the equivalent diameter of the intra-axonal space of myelinated axons, comparing the sham-operated and TBI groups ipsilaterally.}
\label{stats}
\end{figure*}

\clearpage

\renewcommand{\tablename}{Supplementary Table}
\setcounter{table}{0}

\begin{table*}[t]
    \centering
    \caption{Description of the parameters of the segmentation pipeline.}    
    \label{Parameters}    
    \begin{tabularx}{\textwidth}{l|cXcc}
    {\cellcolor[rgb]{0.937,0.937,0.937}}\textbf{Parameter} & 
    {\cellcolor[rgb]{0.937,0.937,0.937}}\textbf{Section} & {\cellcolor[rgb]{0.937,0.937,0.937}}\textbf{Description} &
    {\cellcolor[rgb]{0.937,0.937,0.937}}\textbf{Default} &{\cellcolor[rgb]{0.937,0.937,0.937}}\textbf{Fixed}\\ \hline\hline
    
    $\sigma$ & 2.1.2 & STD for calculating Gaussian smoothed image & 2 & Yes \\ \hline
    $\sigma$ & 2.1.2 & STD for calculating Laplacian of Gaussian image & 0.5 & Yes \\ \hline
    $\sigma$ & 2.1.2 & STD for calculating gradient magnitude image & 2 & Yes \\ \hline
    $\sigma$ & 2.1.2 & STD for calculating difference of Gaussian image & 5 & Yes \\ \hline
    \multirow{2}{*}{H}& \multirow{2}{*}{2.1.3} & H-maxima transform suppresses all local maxima whose height is less than H & \multirow{2}{*}{2} & \multirow{2}{*}{Yes} \\ \hline
    
    \multirow{2}{*}{Similarity threshold} & \multirow{2}{*}{3.2} & Similarity threshold in BVG to determine whether a voxel can be appended to the growing region. & \multirow{2}{*}{0.1} & \multirow{2}{*}{No} \\ \hline
    
    Min volume & 3.2 & In BVG to exclude volumes below this threshold & 100 voxels & No \\ \hline
    Max volume & 3.2 & In BVG to exclude volumes above this threshold & 1,000,000 voxels & No \\ \hline
    
    \multirow{2}{*}{Length threshold} & \multirow{2}{*}{3.4} & In morphology analysis to exclude myelinated axons that are shorter than this threshold & \multirow{2}{*}{4 $\mu m$} & \multirow{2}{*}{Yes} \\ \hline
    
    \multirow{2}{*}{Myelin min volume} & & In semantic segmentation of myelin to exclude volumes under this threshold & \multirow{2}{*}{500 voxels} & \multirow{2}{*}{No} \\ \hline
    
    
    
    \end{tabularx}
\end{table*}

\end{document}